\documentclass[10pt,a4paper,english,aps]{revtex4}
\usepackage[T1]{fontenc}
\usepackage[latin1]{inputenc}
\usepackage{babel}
\usepackage{graphics}

\makeatletter

\providecommand{\LyX}{L\kern-.1667em\lower.25em\hbox{Y}\kern-.125emX\@}

\usepackage[T1]{fontenc}
\usepackage[latin1]{inputenc}
\usepackage{babel}
\usepackage{graphics}

\makeatletter

\usepackage[T1]{fontenc}
\usepackage[latin1]{inputenc}
\usepackage{babel}
\usepackage{graphics}



\makeatother

\makeatother
\begin{document}

\title{Constraints on the interaction and self-interaction of dark energy
from cosmic microwave background}

\author{Luca Amendola, Claudia Quercellini, Domenico Tocchini-Valentini \&
Alessandro Pasqui}

\address{INAF - Osservatorio Astronomico di Roma, \\
Via Frascati 33, 00040 Monte Porzio Catone - Italia}

\begin{abstract}
It is well-known that even high quality cosmic microwave background
(CMB) observations are not sufficient on their own to determine the
equation of state of the dark energy, due to the effect of the so-called
geometric degeneracy at large multipoles and the cosmic variance at
small ones. In contrast, we find that CMB data can put tight constraints
on another fundamental property of the dark energy, namely its coupling
to dark matter. We compare the current high-resolution CMB data to
models of dark energy characterized by an inverse power law or exponential
potential and by the coupling to dark matter. We determine the curve
of degeneracy between the dark energy equation of state and the dimensionless
Hubble parameter \( h \) and show that even an independent perfect
determination of \( h \) may be insufficient to distinguish dark
energy from a pure cosmological constant with the current dataset.
On the other hand, we find that the interaction with dark matter is
firmly bounded, regardless of the potential. In terms of the dimensionless
ratio \( \beta  \) of the dark energy interaction to gravity, we
find \( \beta <0.16 \) (95\% c.l.). This implies that the effective
equation of state between equivalence and tracking has been close
to the pure matter equation of state within 1\% and that scalar gravity
is at least 40 times weaker than tensor gravity. Further, we show
that an experiment limited by cosmic variance only, like the Planck
mission, can put an upper bound \( \beta  \) < 0.05 (95\% c.l.).
This shows that CMB observations have a strong potentiality not only
as a test of cosmic kinematics but also as a gravitational probe.
\end{abstract}
\maketitle

\section{Introduction}

Four years after the first observational hints about the existence
of a dominant component of unclustered matter with negative pressure
(Riess et al. 1998, Perlmutter et al. 1999), the so-called dark energy
or quintessence (Wetterich 1988; Ratra \& Peebles 1988; Frieman et
al. 1995; Caldwell et al. 1998), there are still very few indications
as to its nature. The main reason, perhaps, is that we lack any specific
theoretical suggestion on the properties of the dark energy, i.e.
on its self-interaction potential and on how it interacts with the
other cosmological components. At this stage, all we can do is to
explore a wide range of phenomenological models specified by the potential
and the coupling to the other fields in order to provide an overall
best fit to the current data.

Several works have tried to constrain dark energy models with the
recent high resolution CMB data (Netterfield et al. 2001, Lee et al.
2001, Halverson et al. 2001) leading in some cases to interesting
but conflicting results. In Amendola (2001) the coupling dimensionless
parameter \( \beta  \) that measures the strength of the interaction
of dark energy to dark matter with respect to the gravitational interaction
was constrained to be smaller than 0.1 roughly, adopting an exponential
potential. The slope of the potential, on the other hand, was found
to be essentially unconstrained by the data. Baccigalupi et al. (2002)
found that, among power-law potentials \( V\sim \phi ^{-\alpha } \)
and fixing the Hubble constant to \( h=0.65 \) (in units of \( 100 \)
km/sec/Mpc), values of \( \alpha  \) around unity are favored while
the case of pure cosmological constant, \( \alpha =0 \), was less
likely by a factor of five roughly. Corasaniti \& Copeland (2001),
on the other hand, have shown that the pure cosmological constant
gives the best fit to the data, particularly for as concerns the position
of the acoustic peaks. Bean \& Melchiorri (2002) also conclude that
the pure cosmological constant gives the best fit to the CMB data
when the prior on the Hubble constant is broadened to \( h=0.72\pm 0.08 \)
. All the papers above, and the present one, assume a flat spatial
curvature.

In this paper we extend the previous studies in two respects. First,
we aim at constraining not only the dark energy self-interaction (i.e.
its potential) but also its interaction with matter. To our knowledge,
this is the first work that simultaneously constrain the dark energy
potential and coupling. Second, we put only very weak restrictions
on the cosmological parameters \( h \), \( \Omega _{b,c} \) (the
density parameters of baryons and cold dark matter, respectively),
and \( n_{s} \) (the slope of the scalar perturbations). The conflicting
results above are in fact due mostly to different priors. 

These generalizations allow us to formulate the main conclusion of
the paper: we find that the present CMB data are capable to put a
strong constraint on the coupling but not on the scalar field potential
or equation of state. In fact, the degeneracy between \( h \) and
\( w_{\phi } \), the dark energy equation of state (Huey et al. 1999)
almost erases the sensitivity of the CMB to the dark energy potential.
In sharp contrast, as it will be shown below, the CMB spectra are
very sensitive to the dark energy coupling, since the latter determines
the equation of state for a long stage after equivalence and there
are no strong degeneracies with the other cosmological parameters.

\section{Dark-dark coupling}

Let us consider two components, a scalar field \( \phi  \) and CDM,
described by the energy-momentum tensors \( T_{\mu \nu (\phi )} \)
and \( T_{\mu \nu (c)} \), respectively. General covariance requires
the conservation of their sum, so that it is possible to consider
an interaction between dark energy and dark matter such that \begin{eqnarray}
T_{\nu (\phi );\mu }^{\mu } & = & CT_{(c)}\phi _{;\nu },\nonumber \\
T_{\nu (c);\mu }^{\mu } & = & -CT_{(c)}\phi _{;\nu }.\label{coup1} \\
 & 
\end{eqnarray}
 Such a coupling can be obtained from the conformal transformation
of a Brans-Dicke gravity (see e.g. Amendola 1999) and it has been
considered several times in literature starting from Wetterich (1988,1995)
and Wands et al. (1993). It has been discussed in the context of dark
energy models in Amendola (1999,2000) and in Wands \& Holden (2000),
Chimento et al. (2000), Billyard \& Coley (2000), Chiba (2001), Albrecht
et al. (2001), Esposito-Farese \& Polarsky (2001). In its conformally
related Brans-Dicke form has been studied by Uzan (1999), Chiba (1999),
Chen \& Kamionkowsky (1999), Baccigalupi et al. (2000), Sen \& Sen
(2001). Theoretical motivations in superstring models and in brane
cosmology have been proposed recently in Gasperini, Piazza \& Veneziano
(2002) and Pietroni (2002). 

In the flat conformal FRW metric \( ds^{2}=a^{2}(-d\tau ^{2}+\delta _{ij}dx^{i}dx^{j}) \)
the scalar field and dark matter conservation equations are \begin{eqnarray}
\ddot{\phi }+2H\dot{\phi }+a^{2}U_{,\phi } & = & C\rho _{c}a^{2},\\
\dot{\rho }_{c}+3H\rho _{c} & = & -C\rho _{c}\dot{\phi }\label{kg} 
\end{eqnarray}
 (dots refer to conformal time) where \( H=\dot{a}/a \). Suppose
now that the potential \( U(\phi ) \) is specified by the following
relation \begin{equation}
U^{\prime }=BU^{N}
\end{equation}
 (the prime refers to derivation with respect to \( \phi  \)) which
includes power laws, exponential potentials and a pure cosmological
constant. In the case of power law potential \( U=A\phi ^{-\alpha } \)
we have\begin{equation}
N=(1+\alpha )/\alpha 
\end{equation}
 and \( B=-\alpha A^{(-1/\alpha )}, \) while for the exponential
potential \( N=1 \). We consider only the range \( N\geq 1 \) since
for negative \( \alpha  \) there are no asymptotically accelerating
models; for \( N\rightarrow \infty  \) we recover the pure cosmological
constant. Assuming that the baryons are not directly coupled to the
dark energy (otherwise local gravity experiment would reveal a fifth
force, see Damour et al. 1990) and that the radiation as well is uncoupled
(as it occurs if the coupling is derived by a Brans-Dicke Lagrangian,
see e.g. Amendola 1999), the system of one Einstein equation and four
conservation equations (for radiation, \( \gamma  \), baryons, \( b \),
CDM , \( c \), and scalar field) can be conveniently written introducing
the following five variables that generalize Copeland et al. (1997):
\begin{equation}
x=\frac{\kappa }{H}\frac{\dot{\phi }}{\sqrt{6}},\quad y=\frac{\kappa a}{H}\sqrt{\frac{U}{3}},\quad z=\frac{\kappa a}{H}\sqrt{\frac{\rho _{\gamma }}{3},}\quad v=\frac{\kappa a}{H}\sqrt{\frac{\rho _{b}}{3},}\quad w=\frac{H}{a}
\end{equation}
 where \( \kappa ^{2}=8\pi G \) (notice that \( w=d\log a/dt \)
i.e. the usual Hubble constant). Notice that \( x^{2},y^{2},z^{2}, \)
etc. correspond to the density parameter of each component. In terms
of the independent variable \( \log a \) we have then the system:\begin{eqnarray}
x^{\prime } & = & \left( \frac{z^{\prime }}{z}-1\right) x-\mu y^{2N}w^{2N-2}+\beta (1-x^{2}-y^{2}-z^{2}-v^{2}),\nonumber \\
y^{\prime } & = & \mu xy^{2N-1}w^{2N-2}+y\left( 2+\frac{z^{\prime }}{z}\right) ,\nonumber \\
z^{\prime } & = & -\frac{z}{2}\left( 1-3x^{2}+3y^{2}-z^{2}\right) ,\nonumber \\
v^{\prime } & = & -\frac{v}{2}\left( -3x^{2}+3y^{2}-z^{2}\right) \nonumber \\
w^{\prime } & = & -\frac{w}{2}\left( 3+3x^{2}-3y^{2}+z^{2}\right) \label{gensyst} 
\end{eqnarray}
 where \begin{equation}
\beta =C\sqrt{\frac{3}{2\kappa ^{2}}},\quad \mu =3^{N}\kappa ^{1-2N}\frac{B}{\sqrt{6}}.
\end{equation}
The dimensionless constant \( \beta ^{2} \) can be seen as the ratio
of the dark energy-dark matter interaction with respect to gravity.
It can be shown in fact (Damour \& Nordtvedt 1993, Wetterich 1995)
that the force acting between dark matter particles can be described
in the Newtonian limit as a renormalized Newton's constant \( \hat{G}=G(1+4\beta ^{2}/3) \).
The effect of this interaction on structure growth is discussed in
Amendola \& Tocchini-Valentini (2002).

\section{Critical points and tracking solutions}

The system (\ref{gensyst}) includes several qualitatively different
behaviors, already discussed in Amendola (2000), Tocchini-Valentini
\& Amendola (2002). However, for the range of values that are of cosmological
interest, the system passes through three distinct phases after equivalence.

\( \phi  \)MDE. Immediately after equivalence, the system enters
a matter dominated epoch with a non-negligible \( \phi  \) contribution,
that we denote \( \phi  \)MDE, in which the dark energy potential
density is negligible while the kinetic energy density parameter \( \Omega _{K\phi }=x^{2} \)
of the scalar field gives a constant contribution to the total density.
As will be shown in the following, the existence of such an epoch
is crucial for the constraints that we will be able to put on the
coupling. Putting \( y=0 \) and neglecting radiation and baryons
we obtain the simplified system\begin{eqnarray}
x^{\prime } & = & -\frac{3}{2}x\left( 1-x^{2}\right) +\beta (1-x^{2}),\nonumber \\
y^{\prime } & = & 0.\label{sistphi} \\
 & \nonumber \label{simpsyst} 
\end{eqnarray}
 The point \( y=0 \), \( x=2\beta /3 \) is a saddle point solution
of the system (\ref{sistphi}) for any \( N \) and for \( |\beta |<\sqrt{3}/2 \).
The system stays on the \( \phi  \)MDE solution until \( y \) starts
growing. Along this solution the scale factor expands slower than
a pure MDE, i.e. as\begin{equation}
a\sim t^{4/(6+4\beta ^{2})}.
\end{equation}
 Since \( y=0 \) on the \( \phi  \)MDE, its existence is independent
of the potential, although it has to verified for each potential whether
it is a saddle.

\emph{Tracking trajectories.} Let us now neglect baryons and radiation
and put \( \beta =0 \). The tracking solutions found in Steinhardt
et al. (1999) assume \( y/x=p \) and \( y^{2N-1}w^{2N-2}=q \) where
\( p,q \) are two motion integrals. In the limit \( y^{2},x^{2}\ll 1 \)
it is easy to show that \( p'=q'=0 \) if\begin{eqnarray}
p^{2} & = & 4N-3\label{pn} \\
q & = & -\frac{3p}{2\mu (2N-1)}
\end{eqnarray}
 The same tracking behavior remains a good approximation for small
\( \beta  \). For \( N=1 \) the tracking solution becomes actually
a global attractor of the dynamical system, see below. In the other
cases, the tracking interpolates between the \( \phi  \)MDE and the
global attractor.

\emph{Global attractors}. In the system (\ref{gensyst}) for \( N\not =1 \)
there exists only the attractor \( x=0,y=1 \) on which the dark energy
completely accounts for the matter content. For \( N=1 \) the phase
space is much richer, and there are several possible global attractors,
only two of which accelerated (Amendola 2000). One, for \( \mu >(-\beta +\sqrt{18+\beta ^{2}})/2 \)
, presents a constant non-zero \( \Omega _{c} \): this attractor
actually coincides with the tracking solutions and in fact realizes
the condition \( p=1 \) and \( y=-3/(2\mu ) \) as requested by Eq.
(\ref{pn}) for \( \beta =0 \). This {}``stationary attractor{}''
can be accelerated if \( \beta >2\mu  \) and could solve the coincidence
problem since \( \Omega _{c}\propto \Omega _{\phi } \); it has been
discussed in detail in Amendola \& Tocchini-Valentini (2001, 2002).
The other accelerated attractor occurs when \( \mu <(-\beta +\sqrt{18+\beta ^{2}})/2 \):
in this case there are no tracking solutions and the global attractor
reduces to \( x=0,y=1 \) as for \( N\not =1 \). These solutions
have already been compared to CMB for \( N=1 \) in Amendola (2001).
The inclusion of the baryons modifies the considerations above but,
as long as they are much smaller than the other components, the qualitative
behavior of the system remains the same. It is to be noticed that
the final attractor, on which the dark energy dominates completely
the cosmic fluid, is yet to be reached, and therefore the existence
of an accelerated epoch at the present depends mostly on the tracking.

The existence of the \( \phi  \)MDE saddle and of the tracking solutions
is crucial for our analysis. In fact, these two epochs guarantee that
the equation of state of the scalar field is piece-wise constant through
essentially all the post-equivalence epoch. In the \( \phi  \)MDE
phase the effective parameter of state \( w_{e}=p_{tot}/\rho _{tot}+1 \)
and the field equation of state \( w_{\phi }=p_{\phi }/\rho _{\phi }+1 \)
are\begin{equation}
\label{phimde}
w_{e}=1+\frac{4}{9}\beta ^{2},\quad w_{\phi }=2
\end{equation}
 while during the tracking phase \begin{equation}
\label{track}
w_{e}\approx 1,\quad w_{\phi }\approx \frac{2}{1+p^{2}}=\frac{1}{2N-1}
\end{equation}
 (the last relation is approximated and it is actually only an upper
bound to the present \( w_{\phi } \); it is more precise if \( w_{\phi } \)
is identified with the average equation of state after \( \phi  \)MDE
rather than the present equation of state). Therefore, the cosmic
evolution depends on \( \beta  \) alone during the \( \phi  \)MDE,
and on \( N \) alone during the tracking. Since the position of the
acoustic peaks is related to the equation of state through the angular
diameter distance, it appears that the CMB is able to put direct constraints
on both \( \beta  \) and \( N \). In Fig. 1 we show a typical trajectory
that presents in sequence the three epochs discussed above.

\begin{figure}
{\centering \resizebox*{!}{14cm}{\includegraphics{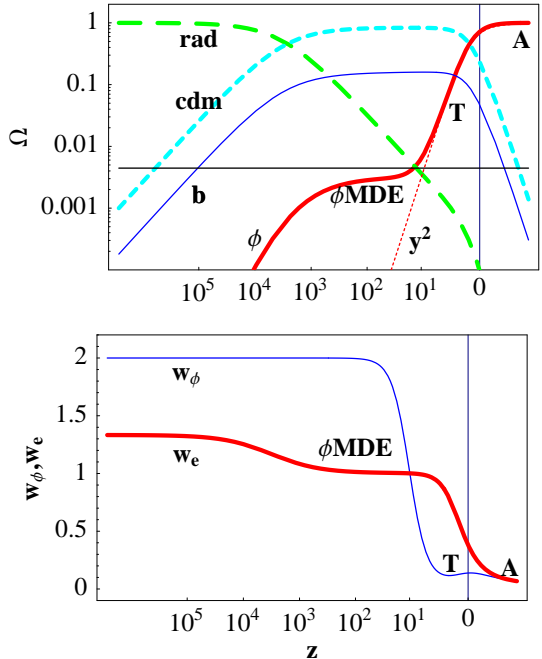}} \par}

\caption{Numerical solutions of the system (\ref{gensyst}) for \protect\( N=2,\beta =0.1,\omega _{c}=0.1,\omega _{b}=0.02,h=0.65\protect \)
plotted against the redshift. \emph{Upper panel}. Long dashed line:
radiation; short dashed line: CDM; unbroken thin line: baryons; unbroken
thick line: scalar field; dotted line: scalar field potential energy.
The horizontal thin line marks the kinetic energy density of \protect\( \phi \protect \)MDE,
reached just after equivalence. \emph{Lower panel.} Thick line: effective
equation of state; thin line: dark energy equation of state. The labels
mark the \protect\( \phi \protect \)MDE, the tracking (T) and the
final attractor (A).}
\end{figure}

There is however a crucial difference between the \( \phi  \)MDE
and a tracking for what concerns here: while the present dark energy
equation of state, set by the tracking, is degenerated with \( h \)
for as concerns the CMB spectrum (see e.g. Huey et al. 1999; Bean
\& Melchiorri 2002), the equation of state during the \( \phi  \)MDE
is not. In fact, the angular diameter distance to the last scattering
surface \( d_{A} \) is degenerate along lines \( h(w_{\phi }) \)
for which\begin{equation}
\label{deg}
d_{A}\sim \int _{a_{dec}}^{1}da\left[ \omega _{c}a+(h^{2}-\omega _{c})a^{4-3w_{\phi }}\right] ^{-1/2}=const
\end{equation}
 where \( \omega _{c}\equiv \Omega _{c}h^{2} \) (although this is
exact only for \( \beta =0 \) it remains a good approximation even
for small non-zero values). Assuming a strong prior on \( h \) a
peak emerges in the likelihood for \( N \) but then the result is
clearly prior-dependent. The same holds true for models in which the
equation of state is slowly varying (see e.g. Huey et al. 1999, Doran
et al. 2001). On the other hand, the fact that \( \Omega _{\phi }\not =0 \)
at decoupling in coupled models implies that the effects of the coupling
on the CMB are not due solely to the angular diameter distance, and
therefore the geometric degeneracy can be broken. This is shown in
Fig. 2 in which \( C_{\ell } \) spectra for various values of \( \beta  \)
(all other parameters being equal) are shown: the spectra change both
in amplitude and in peak's position.

\begin{figure}
{\centering \resizebox*{!}{12cm}{\includegraphics{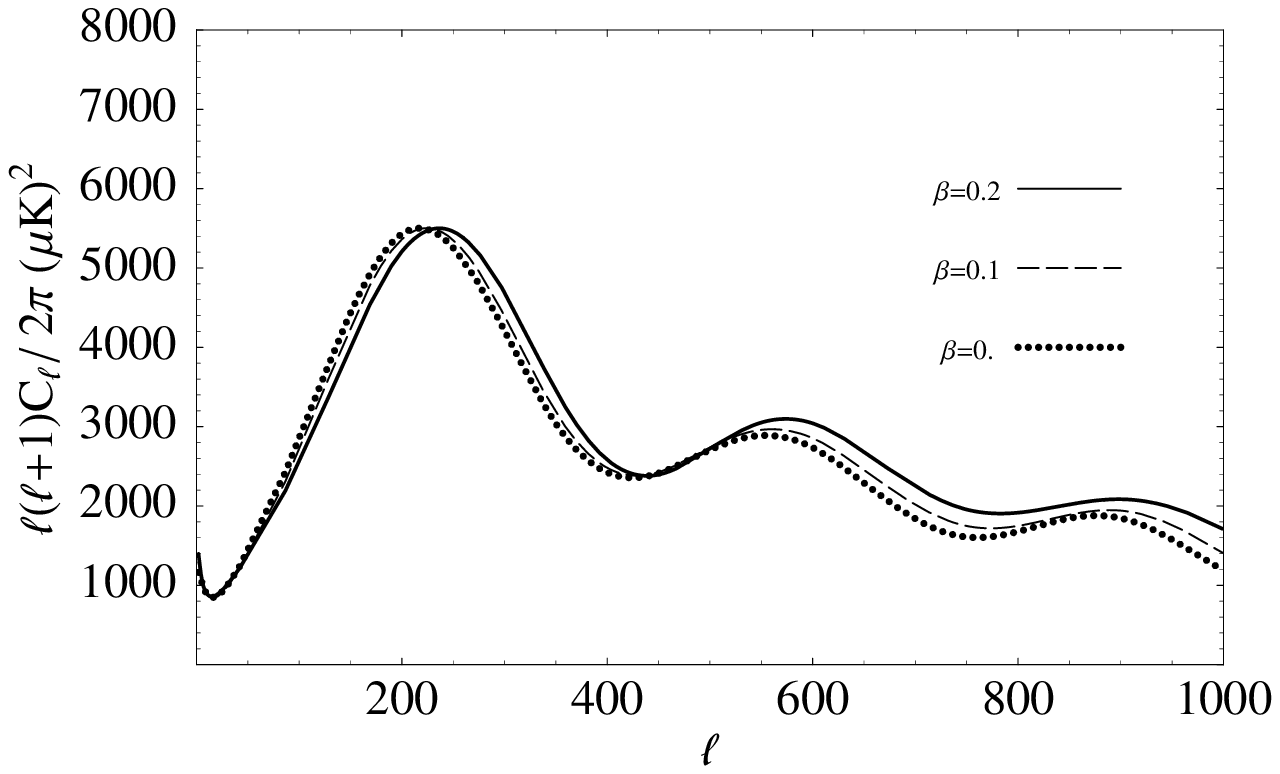}} \par}

\caption{CMB spectra for increasing values of the coupling constant (the other
parameters are \protect\( N=1.5,h=0.65,\omega _{b}=0.01,\omega _{c}=0.1\protect \)).
Notice that the peaks not only move to the right but also change in
height.}
\end{figure}

\section{CMB constraints}

Our theoretical model depends on three scalar field parameters and
four cosmological parameters: \begin{equation}
\beta ,\mu ,N,n_{s},h,\omega _{b},\omega _{c}
\end{equation}
 where \( \omega _{b}=\Omega _{b}h^{2} \) and \( \omega _{c}=\Omega _{c}h^{2} \).
We calculate the \( C_{\ell } \) spectra by a modified CMBFAST (Seljak
\& Zaldarriaga 1996) code that includes the full set of coupled perturbation
equations (see Amendola 2000 and Amendola \& Tocchini-Valentini 2001).
The other parameters are set as follows: \( T_{cmb}=2.726K, \) \( Y_{He}=0.24,N_{\nu }=3.04,\tau _{c}=0. \)
In the analysis of Netterfield et al. (2002) \( \tau _{c} \), the
optical depth to Thomson scattering, was also included in the general
likelihood but, in the flat case, was found to be compatible with
zero and to have only a minor effect on the other parameters. We did
not include the cross-correlation between bandpowers and pointing
and beam corrections because they are not available for all the experiments;
we adopted a pure log-normal likelihood. We used as dataset the COBE
data analyzed in Bond et al. (2000), and the high resolution data
of Boomerang (Netterfield et al. 2002), Maxima (Lee et al. 2002),
and DASI (Halverson et al. 2002). The overall amplitude and the calibration
errors of Boomerang (10\%) and of Maxima and DASI (4\%) have been
integrated out analytically. The theoretical spectra have been binned
as the experimental ones.

In Amendola (2001) we have shown that the dynamics of the system is
insensitive to the sign of \( \beta  \), since \( \phi  \)MDE, tracking
and the final accelerated epoch do not depend on it. We will consider
only \( \beta \geq 0 \).

The tracking trajectories are selected for each \( N,\beta  \) by
fixing the initial conditions on \( x,y \), and then varying \( \mu  \)
until the present values of \( h,\omega _{b},\omega _{c} \) are found.
Therefore, the parameter \( \mu  \) is not a free parameter, but
is fixed in function of the others. In fact, if there is tracking,
all trajectories with equal \( N,\beta ,h,\omega _{b},\omega _{c} \)
always had the same equation of state and matter densities, and therefore
generate a similar CMB spectrum, so that it is sufficient to select
one of them. There are in general initial conditions that do not lead
to tracking (but still undergo \( \phi  \)MDE). These will be discussed
in another work.

In order to compare with the previous analyses we assume uniform priors
with the parameters confined in the range \( \beta \in (0,0.3), \)
\( \quad N\in (1,8.5),\quad  \)\ \( n_{s}\in (0.7,1.3), \) \( h\in (0.45,0.9), \)
\( \quad \omega _{b}\in (0.005,0.05),\quad  \)\ \( \omega _{c}\in (0.01,0.3) \)
. The same age constraints (\( >10 \) Gyr) used in most previous
analyses is adopted here. A grid of \( \sim 10,000 \) multipole CMB
spectra \( C_{\ell } \) is used as a database over which we interpolate
to produce the likelihood function. Notice that we cannot use here
the rescaling of the high-\( \ell  \) spectra adopted in Bean \&
Melchiorri (2002) to speed up calculations, due to the non-vanishing
scalar field energy density during the \( \phi  \)MDE.

\section{Likelihood results}

In Fig. 3 we show the likelihood curves for \( N,h \) marginalizing
over the remaining parameters. The contours of the likelihood plot
follow the expected degeneracy of the angular diameter distance. The
likelihood is almost degenerated along the dotted curve \begin{equation}
h=0.94-0.56w_{\phi }+0.03w_{\phi }^{2}
\end{equation}
 In the plot we also show the expected degeneracy curve from Eq. (\ref{deg})
assuming \( \omega _{c}=0.1 \). The residual deviation from the expected
degeneracy line is due to the age prior (>10 Gyr) that favors small
\( h \) values. Notice that for \( h>0.75 \) no upper limit to \( N \)
can be given with the current CMB data no matter how precise \( h \)
is, and that only assuming \( h<0.65 \) it becomes possible to exclude
at \( 95\% \) c.l. the pure cosmological constant. In Fig. 4 we show
the likelihood for all parameters, marginalizing in turn over the
others. Notice that the likelihood for \( N \) flattens for \( N>1.2 \).
We find the following constraints at 95\% c.l.:\begin{equation}
N>1.5,\quad \beta <0.16
\end{equation}
 (the limit on \( N \) corresponds to \( \alpha <2 \)). The limit
on \( N \) is however only a formal one: the likelihood never vanishes
in the definition domain and even \( N=1 \) (the exponential potential)
is only a factor of five less likely than the peak and certainly cannot
be excluded on this basis. Moreover, the lower bound on \( N \) is
prior-dependent: allowing smaller values of \( h \) it would weaken.
Clearly, if we adopt narrow priors on \( h \), we can indeed obtain
more stringent bounds on \( N \), as shown in the same Fig. 4. In
contrast, the constraint on \( \beta  \) does not depend sensibly
on the prior on \( h \) and the likelihood for \( \beta  \) does
vanish at large \( \beta  \). In place of \( N \) and \( \beta  \)
we can use as well the equation of state during tracking and during
\( \phi  \)MDE, respectively, as likelihood variables, using Eqs.
(\ref{phimde}) and (\ref{track}). Then we obtain at the 95\% c.l.\begin{equation}
\label{cons}
w_{\phi (tracking)}<0.8,\quad 1<w_{e(\phi MDE)}<1.01
\end{equation}
 This shows that the effective equation of state during \( \phi  \)MDE,
i.e. between equivalence and tracking, is close to unity (as in a
pure matter dominated epoch) to within one per cent. The striking
difference between the level of the two constraints in (\ref{cons})
well illustrates the main point of this paper: the CMB is much more
sensitive to the dark energy coupling than to its potential. 

The other parameters are (we give here for simplicity the mean and
the one sigma error, while the limit is at the 95\% c.l.) \begin{equation}
n_{s}=0.99\pm 0.05,\quad \omega _{b}=0.023\pm 0.004,\quad \omega _{c}=0.092\pm 0.02,\quad h>0.62
\end{equation}
The total dark energy density turns out to be \( \Omega _{\phi }>0.60 \)
(95\% c.l.). It appears that the limits on the cosmological parameters
\( n_{s,}\omega _{b},\omega _{c} \) are almost independent of \( \beta  \),
while a non-zero \( \beta  \) favors higher \( h \) (Fig. 4, panel
\( d \)). It is interesting to compare with the current constraints
from the Hubble diagram of the supernovae Ia, where one obtains quite
a stronger bound on the equation of state, \( w_{\phi }<0.4 \) or
\( N>1.75 \) at the same c.l.. 

Finally, in Fig. 4 (panel \emph{b)} we plot the likelihood for \( \beta  \)
that an experiment with no calibration uncertainty and limited only
by cosmic variance, like the Planck mission (de Zotti et al. 1999),
can achieve. We find \( \beta <0.05 \) (95\% c.l.) (using the specifications
of the MAP satellite we find \( \beta <0.1 \)).

\begin{figure}
{\centering \resizebox*{!}{12cm}{\includegraphics{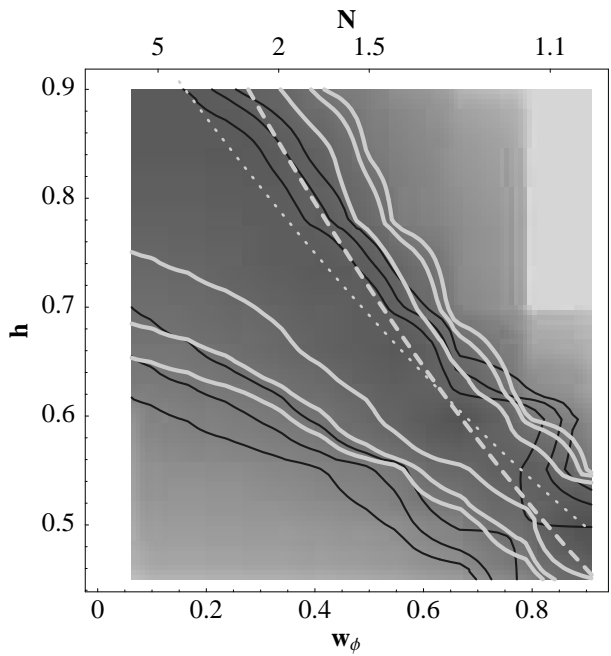}} \par}

\caption{Likelihood contour plots in the space \protect\protect\protect\( N,h\protect \protect \protect \)
marginalizing over the other parameters at the 68,90 and 95\% c.l..
The white thick lines refer to the coupled case, the black curves
to the uncoupled case. The dotted line is the likelihood degeneracy
curve, the dashed line is the expected degeneracy curve. Notice that
only fixing \protect\protect\protect\( h\protect \protect \protect \)
smaller than 0.65 it would be possible to exclude the cosmological
constant at 95\% c.l..}
\end{figure}

\begin{figure}
{\centering \resizebox*{12cm}{!}{\includegraphics{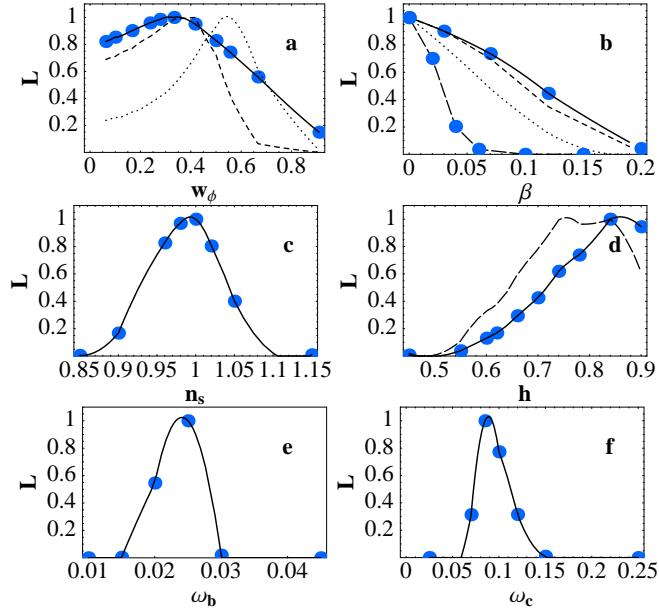}} \par}

\caption{Marginalized likelihood for tracking trajectories. The equation of
state in panel \protect\protect\protect\( a\protect \protect \protect \)
is \protect\protect\protect\( w_{\phi }=1/(2N-1)\protect \protect \protect \).
In panels a and b we plot as a dotted line the likelihood assuming
\protect\protect\protect\( h=0.65\pm 0.05\protect \protect \protect \)
and as dashed line \protect\protect\protect\( h=0.75\pm 0.05\protect \protect \protect \)
(gaussian prior). In panel \protect\( b\protect \) the long-dashed
line is the Planck-like likelihood. In panel \protect\( d\protect \)
the long-dashed line is for \protect\( \beta =0\protect \).}
\end{figure}

\section{Conclusions}

The main result of this paper is that CMB observations are a powerful
gravitational probe able to constrain scalar gravity. In fact, a dark
matter-dark energy interaction would obviously escape any local gravity
experiment: cosmological observations like the CMB are then the only
way to observe such a phenomenon. Since observations require the baryons
to be decoupled from dark energy (or coupled much more weakly than
dark matter), the search for a non-zero \( \beta  \) is in fact also
a test of the equivalence principle. We found that current CMB data
are capable to put an interesting upper bound to the dark matter -
dark energy coupling: \[
\beta <0.16\]
 (95\% c.l.) regardless of the potential (within the class we considered).
This implies that the scalar gravity is at least \( 1/\beta ^{2}\approx 40 \)
times weaker than ordinary tensor gravity. As shown in Amendola (1999),
the limit on \( \beta  \) can be restated as a limit on the constant
\( \xi  \) of the non-minimally coupled gravity, \( \xi <0.017 \).
An experiment like the Planck mission can lower the upper bound to
\( \beta  \) to 0.05: scalar gravity would be in this case at least
\( 400 \) times weaker than ordinary tensor gravity. This limit is
comparable to those that local gravity experiments impose on the scalar
gravity coupling to baryons, \( \beta _{baryons}^{2}<10^{-3} \) (see
e.g. Groom et al. 2000).

In contrast, CMB data, on their own, cannot put \emph{any} firm limit
to the dark energy potential, unless a narrow prior on \( h \) is
adopted: e.g. \( h=0.65\pm 0.05 \) gives \( w_{\phi }=0.55\pm 0.2 \)
while \( h=0.75\pm 0.05 \) gives \( w_{\phi }=0.35\pm 0.2 \) (assuming
gaussian priors and marginalizing as usual over all the other parameters,
including the coupling).

In a subsequent work we will consider also the trajectories that do
not undergo tracking: in this case the field \( \phi  \) is frozen
until recently, and the trajectories are indistinguishable from a
pure cosmological constant, except for the existence of a \( \phi  \)MDE
if \( \beta \not =0 \). As expected, we will show that even for off-tracking
trajectories the CMB data constrain \( \beta  \) but not the potential.
This shows that the constraint on \( \beta  \) is independent both
of the potential and of the initial conditions.

\end{document}